\documentclass[reprint,aps,floats,floatfix,amsmath,amssymb,amsfonts,nofootinbib,longbibliography,superscriptaddress]{revtex4-1}

\usepackage{graphicx}
\usepackage{dcolumn}
\usepackage{bm}
\usepackage{tensor} 
\usepackage{natbib}
\usepackage{subcaption}

\usepackage[hyperindex,colorlinks]{hyperref}

\begin{document}

\title{Nonconservative unimodular gravity: a viable cosmological scenario}
\author{J\'ulio C. Fabris}
\email{julio.fabris@cosmo-ufes.org}%
\affiliation{%
N\'ucleo Cosmo-ufes \& Departamento de F\'isica,  Universidade Federal do Esp\'irito Santo (UFES)\\
Av. Fernando Ferrari, 540, CEP 29.075-910, Vit\'oria, ES, Brazil.}%
\affiliation{%
National Research Nuclear University MEPhI, Kashirskoe sh. 31, Moscow 115409, Russia}%

\author{Marcelo H. Alvarenga}
\email{marcelo.alvarenga@edu.ufes.br}
\affiliation{%
N\'ucleo Cosmo-ufes \& Departamento de F\'isica,  Universidade Federal do Esp\'irito Santo (UFES)\\
Av. Fernando Ferrari, 540, CEP 29.075-910, Vit\'oria, ES, Brazil.}%
\author{Mahamadou Hamani-Daouda}%
\email{daoudah77@gmail.com}

\affiliation{Département de Physique - Université de Niamey, Niamey, Niger.}%

\author{Hermano Velten}%
\email{hermano.velten@ufop.edu.br}
\affiliation{%
Departamento de F\'isica, Universidade Federal de Ouro Preto (UFOP), Campus Universit\'ario Morro do Cruzeiro, 35.400-000, Ouro Preto, Brazil}%
\date{\today}

\begin{abstract}
In this work we review the issue of imposing the conservation of the energy-momentum tensor as a necessary condition to recover the equivalence between the unimodular gravity and General Relativity (GR) equipped with a cosmological constant. This procedure is usually interpreted as an {\it ad hoc} imposition on the unimodular theory's structure. Whereas the consequences of avoiding the conservation of the total energy-momentum tensor has been already introduced in the literature, 
it has been not widely explored so far. We study an expanding universe sourced by a single effective perfect fluid such that the null divergence of its energy-momentum tensor is not imposed. As we shall show, in this scheme, the unimodular theory has its own conservation equation obtained from the Bianchi identities. We explore the evolution of the homogeneous and isotropic expanding background and show that a viable cosmological scenario exists.  Also, we consider scalar perturbations with particular attention given to the gauge issue. We show that contrary to the traditional unimodular theory where the synchronous and longitudinal (newtonian) gauge for cosmological perturbations are not permitted, if the conservation of the energy-momentum is relaxed the scalar perturbations in the synchronous condition survive and present a growing mode behavior. We study therefore a new cosmological scenario in which the dynamics of the universe transits from the radiative phase directly to a accelerated one but allowing thus for structure formation. 
\end{abstract}

\maketitle

\section{Introduction}
Current cosmological data are well described by a general relativistic description of the gravitational interaction sourced by perfect fluids and a cosmological constant $\Lambda$. In the standard cosmological scenario the universe experiences three different dynamical epochs usually named as the radiation, the matter and the dark energy eras. Within this scenario one associates in the latter era the quantum vacuum effects on large scales to $\Lambda$ \cite{Zeldovich:1967gd} but, on the other hand, this mechanism gives rise to the famous ``cosmological constant problem" (CCP) \cite{Weinberg:1988cp,Martin:2012bt}. At the same time, it is well known that unimodular gravity, a gauge fixed version of general relativity introduced by Einstein in 1919 in which $\Lambda$ appears as an integration constant of the field equations, is formally equivalent to the case $GR + \Lambda$. Therefore, a possible route to circumvent the CCP is the adoption of the unimodular gravity since in this scenario one does not have to assume that vacuum energy will have relevant gravitational effects \cite{Ellis:2010uc}. It is then natural to expect that the following question should appear: Is it possible to differentiate between both approaches? This issue has been widely explored in the literature both at the classical \cite{doi:10.1063/1.529283,Alvarez:2007nn,Alvarez:2012px,Jain:2011jc,Jain:2012gc} and quantum levels \cite{Alvarez:2005iy,deBrito:2021pmw,Alvarez:2015sba,Bufalo:2015wda,Percacci:2017fsy,deBrito:2020xhy,Eichhorn:2013xr}. In our viewpoint the study of the cosmological evolution filled with perfect fluids is sufficient to differentiate between both approaches. In this work we will explore in more details this possibility (see also \cite{Alvarez:2021cxy}).

One can briefly review the essence of such equivalence by considering the total action of the theory formed by the sum of the gravitational one ${\cal S}_g$ and the matter part ${\cal S}_m$ i.e., ${\cal S}= {\cal S}_g+{\cal S}_m$,
\begin{eqnarray}
{\cal S}_g&=&\int d^4x\biggr\{\sqrt{-g}R - \chi(\sqrt{-g} - \xi)\biggl\},  \\
{\cal S}_m&=&\int d^4 x \sqrt{-g}{\cal L}_m.
\end{eqnarray}

In the above action one can easily identify $\chi$ as a Lagrange multiplier. The unimodular condition (a gauge fixed version of GR) forces the determinant of the metric to obey a specific constraint. Indeed, by varying the total action ${\cal S}$ with respect to $\chi$ one obtains
\begin{eqnarray}
\label{vin-rg-1}
\xi = \sqrt{-g}.
\end{eqnarray}

On the other hand, by varying the total action ${\cal S}$ with respect to the metric $g_{\mu\nu}$ one obtains
\begin{eqnarray}
\label{erg1}
R_{\mu\nu} - \frac{1}{2}g_{\mu\nu}R + \frac{\chi}{2}g_{\mu\nu} = 8\pi GT_{\mu\nu}.
\end{eqnarray}

From the trace of (\ref{erg1})  one obtains a constraining equation for the Lagrange multiplier
\begin{eqnarray}
\chi = \frac{R}{2} + 8\pi G\frac{T}{2},
\end{eqnarray}
which can be inserted again back into (\ref{erg1}) leading to
 \begin{eqnarray}
 \label{erg2}
 R_{\mu\nu} - \frac{1}{4}g_{\mu\nu}R = 8\pi G\biggr(T_{\mu\nu} - \frac{1}{4}g_{\mu\nu}T\biggl).
 \end{eqnarray}
 In principle the above equation looks like a modification of Einstein's equation. But up to this point it would be too naive to state that one has elaborated a new version of the gravitation field equations with potentially testable predictions since the vacuum version of (\ref{erg2}) can be recasted in the same fashion as in GR. 
 
 Now by using the Bianchi identities one obtains the following relation
 \begin{eqnarray}
 \label{brg1}
 \frac{R^{;\nu}}{4} = 8\pi G\biggr({T^{\mu\nu}}_{;\mu} - \frac{1}{4}T^{;\nu}\biggl),
 \end{eqnarray}
that can be seen as a modified conservation law. However, at this point another fundamental principle is usually evoked now: The conservation of energy and momentum. Thus if we impose that the energy-momentum tensor conserves separately, i.e.,
 \begin{eqnarray}
 \label{cons-rg-1}
 {T^{\mu\nu}}_{;\mu} = 0,
 \end{eqnarray}
 we will find out that equation (\ref{brg1}) becomes,
 \begin{eqnarray}
 \label{brg2}
 \frac{R^{;\nu}}{4} = -2 \pi GT^{;\nu}.
 \end{eqnarray}
The above choice is a way to circumvent the fact that in unimodular there are 9 independent equations (differently from GR with 10 independent equations).

But equation (\ref{brg2}) can be integrating leading to,
\begin{eqnarray}
R = - 8\pi GT - 4\Lambda,
\end{eqnarray}
where $\Lambda$ is an integration constant which plays the rôle of a cosmological constant. Indeed, inserting this relation in (\ref{erg1}), we obtain,
\begin{eqnarray}
 \label{erg3}
 R_{\mu\nu} - \frac{1}{2}g_{\mu\nu}R = 8\pi GT_{\mu\nu} + g_{\mu\nu}\Lambda.
 \end{eqnarray}
 This is equivalent to the RG equations with a cosmological constant that appears as a integration constant.
 
Our analysis is focused on the conservation law (\ref{cons-rg-1}). Is it indeed a necessary condition? The answer is, yes. But only if one wants to recover the standard $GR + \Lambda$ scenario. The conservation of $T_{\mu\nu}$ is no longer a consequence of the Bianchi identities but an imposition on the theory's structure \cite{Ellis:2010uc,Ellis:2013uxa}. This issue motivates the following question: What are the consequences of avoiding the conservation of $T_{\mu\nu}$? Indeed, there are multiple examples of nonconservative theories of gravity (see Ref. \cite{Velten:2021xxw} for a review of such theories and \cite{Josset:2016vrq} for physical motivations in the context of dark energy) that can serve as motivation to investigate the case
\begin{eqnarray}
 \label{noncons-rg-1}
 {T^{\mu\nu}}_{;\mu} \neq 0.
 \end{eqnarray}

The proposal of a nonconservative unimodular gravity has already appeared in Ref. \cite{Astorga-Moreno:2019uin} applying it to the description of compact objects.

The purpose of the present analysis is to verify the consequences of not imposing the separate conservation of the energy-momentum tensor to the cosmological arena. Our analysis focus on the gravitational interaction at the classical level and using the background expanding cosmological and its scalar perturbations as the case study. 
 
\section{A new background cosmological model}

Let us turn to the flat, homogeneous and isotropic expanding cosmological background. The so called Friedmann-Lemaitre-Robertson-Walker (FLRW) is given by the metric
\begin{eqnarray}
\label{metric}
ds^2 = N^2dt^2 - a(t)^2(dx^2 + dy^2 + dz^2).
\end{eqnarray}
Then the field equations and the conservation laws are given by the following set (by adopting $N = 1$)
\begin{eqnarray}
\label{ce1}
\dot H &=& - 4\pi G(\rho + p),\\
\label{ce2}
\ddot H + 4H\dot H &=& - 4\pi G[\dot\rho + \dot p + 4H(\rho + p)]. 
\end{eqnarray}
In these equations the expansion rate is given by $H = \dot a/a$ where a dot means derivative with respect to the cosmic time $t$.
In fact, these two equations have the same content: inserting (\ref{ce1}) into (\ref{ce2}) we obtain the identity $0 = 0$. 

Notice that because of the condition
\begin{eqnarray}
g = 1, 
\end{eqnarray}
we should use $N = a^{-3}$. But this {\it correct} lapse function can be restored by choosing a convenient time coordinate. However, in \cite{Gao:2014nia}, $\xi$ is considered as a fixed function of time and so the condition $\xi = \sqrt{-g}$ allows any time gauge through a convenient choice of $\xi$.

Let us define the barred quantity
\begin{equation}
    \bar \rho = \rho + p,
\end{equation} 
which can be interpreted as the enthalpy of the system. Then, equations (\ref{ce1}) and (\ref{ce2}) can be written as,
\begin{eqnarray}
\label{ce1A}
\dot H &=& - 4\pi G\bar\rho,\\
\label{ce2A}
\ddot H + 4H\dot H &=& - 4\pi G(\dot{\bar\rho} + 4H\bar\rho). 
\end{eqnarray}

The system formed by equations (\ref{ce1A}) and (\ref{ce2A}) is underdetermined. Hence, one can suppose any behavior for either the density or for the scale factor. But, we look for a viable cosmological model. In this sense, and in order to have an agreement with observations, one desires an initial radiative dynamical regime and reach asymptotically a de Sitter phase. These requirements will guide us in determining a specific solution. The standard cosmological model requires also a matter dominated phase in order to have structure formation. But, as we will verify later, this requirement is not obligatory in the nonconservative unimodular cosmology.

It is worth noting that the usual radiative solution of GR is also solution here for any $p \neq - \rho$ value
\begin{eqnarray}
H = \frac{1}{2t}, \quad \bar\rho = \bar\rho_0 a^{-4}.
\end{eqnarray}

For $p = - \rho$ we find the usual de Sitter solution, $a \propto e^{\kappa t}$, $\kappa$ a constant (positive or negative).

Is there any other solution? Let us inspect this possibility. We can rewrite (\ref{ce2}) as
\begin{eqnarray}
\frac{d}{dt}\biggr(e^I \dot H\biggl) = - 4\pi G\frac{d}{dt}\biggr(e^I\bar\rho\biggl), \quad I = 4\int Hdt.
\end{eqnarray}
leading to,
\begin{eqnarray}
\dot H = - 4\pi G\bar\rho + ce^{-I}, \quad c = \mbox{constant}.
\end{eqnarray}
However, from equation (\ref{ce1}) one has to impose $c = 0$. This reinforces the fact that the system in incomplete since there is only one equation to determine two functions, namely $H$ and $\rho$. Hence, the system can not be solved if one additional ansatz is introduced. Once more, if the conservation of the energy-momentum tensor is imposed this restriction disappears. Therefore, the conservation of the energy-momentum tensor plays de rôle of the additional constraint equation. But the goal in our analysis here is not to use such ansatz, keeping relation (\ref{brg1}) as it is.

A direct inspection shows that, in order to obtain the necessary features for a viable cosmological model in the context developed so far, it is enough to impose that both sides of (\ref{ce2}) conserve separately. This leads to,
\begin{eqnarray}
\label{cea}
\ddot H + 4H\dot H = 0,\\
\label{ceb}
\dot{\bar\rho} + 4H\bar\rho = 0.
\end{eqnarray}

The above equation (\ref{ceb}) has a simple solution
\begin{eqnarray}
\bar\rho = \bar\rho_0 a^{-4},
\end{eqnarray}
which corresponds to the typical scaling law of the radiative fluid in GR. However, $\bar\rho = (\rho + p)$ and hence the radiative behavior is always obtained independently of the fluid considered. Any perfect fluid energy density will scale according the the radiative behavior. In other words, the physics depends only on the combination $\rho+p$ (see \cite{Alvarez:2021cxy}). This is due to the traceless character of the field equations and also because the energy-momentum tensor is not conserved separately (otherwise we recover the full GR structure). 

Besides the feature discussed above that is valid for any perfect fluid which is similar to a radiative fluid and independent on the adopted pressure, there is another subtle difference. In GR the background evolution equations for a pure radiative fluid are given by,
\begin{eqnarray}
H^2 &=& \frac{8\pi G}{3}\rho_r,\\
2\dot H + 3H^2 &=& - \frac{8\pi G}{3}\rho_r.
\end{eqnarray}
These equations lead to,
\begin{eqnarray}
\dot H + 2H^2 = 0.
\end{eqnarray}

At the same time, it is worth noting that equation (\ref{cea}) can be written as
\begin{eqnarray}
\frac{d}{dt}\biggr(\dot H + 2H^2\biggl) = 0.
\end{eqnarray}
This is equivalent to the result found in Ref. \cite{Daouda:2018kuo} where the constraint condition $R = $ constant had been introduced in order to obtain a closed set of equations. 

Hence, in the nonconservative unimodular cosmology i.e., without adopting a separated energy-momentum tensor conservation, in order to satisfy the requirements described above the expansion rate is determined by,
\begin{eqnarray}
\label{cec}
\dot H + 2H^2 =  \frac{2}{3}\Lambda_{\rm U}, \quad \Lambda_{\rm U}  = \mbox{constant}.
\end{eqnarray}
The integration constant, which we have called $\Lambda_U$, makes the unimodular cosmological scenario essentially identical to the GR radiative model in presence of a cosmological constant. As shall show bellow, it is convenient to introduce the factor $2/3$.

From (\ref{cec}) we have three possibilities:
\begin{eqnarray}
\Lambda_{\rm U} < 0 \quad &\rightarrow& \quad a = a_0\sin^{1/2}\sqrt{-\frac{4 \Lambda_{\rm U}}{3}} t,\\
\Lambda_{\rm U} = 0 \quad &\rightarrow& \quad a = a_0 t^{1/2},\\
\Lambda_{\rm U} > 0 \quad &\rightarrow& \quad a = a_0\sinh^{1/2}\sqrt{ \frac{4 \Lambda_{\rm U}}{3}}t. \label{apositiveLambda}
\end{eqnarray}
These are essentially the same solutions found in Ref. \cite{Daouda:2018kuo}.
The case $\Lambda_{\rm U} = 0$ is identical to the GR radiative model. The solutions corresponding to $\Lambda_{\rm U} \neq 0$ could also be expressed in terms of $\cos$ and $\cosh$ functions, but these possibilities would imply in a negative barred energy density $\bar\rho$, which mounts to a violation of the null energy condition ($\rho + p < 0$).

For all three possible values of $\Lambda_{\rm U}$ the behaviour of the initial phase is similar and coincides with the flat radiative case. The resulting evolution for the $\Lambda \neq 0$ cases are the following. The cosmic dynamics transits from the initial radiative phase to a de Sitter (anti-de Sitter) if $\Lambda_{\rm U} > 0$ ($\Lambda_{\rm U} < 0$). These results indicate the possibility of a transition from the radiative dominated universe to
a de Sitter expansion when $\Lambda_{\rm U}>0$, in contrast to the $\Lambda$CDM model that interpolates a
matter dominate universe and a de Sitter phase.

Let us investigate in more details the background solution obtained with $\Lambda_{\rm U}>0$ since this is potentially the most interesting one.  From Eq. (\ref{apositiveLambda}) the expansion rate and the deceleration parameter read, respectively
\begin{equation}
    H(t)=\sqrt{\frac{\Lambda_{\rm U}}{3}} {\rm Coth} \left[2\sqrt{\frac{\Lambda_{\rm U}}{3}}\, t\right],
    \label{Ht}
\end{equation}

\begin{equation}
    q(t)=-1-\frac{\dot{H}}{H^2}=1-\frac{3}{2}\, {\rm Tanh}^2 \left[2 \sqrt{\frac{\Lambda_{\rm U}}{3}}\, t\right].
\label{q}
\end{equation}

The solution for the deceleration parameter $q(t)$ in (\ref{q}) transits from the asymptotic past value $q(t\rightarrow0) = + 1$ (as in the radiative case) to the de-Sitter expansion in the far future $q(t\rightarrow + \infty) = -1 $. The moment of the transition to the accelerated phase depends uniquely on the value of the constant $\Lambda_{\rm U}$.  
From (\ref{Ht}) and (\ref{q}) we can obtain the following relation for the today's deceleration parameter
\begin{equation}
    q_0=1-\frac{2\Lambda_{\rm U}}{3H^2_0}.
    \label{q0}
\end{equation}
Therefore, depending on the $\Lambda_{\rm U}$ value the universe evolution can experience a current accelerated expansion phase since $\Lambda_{\rm U} > 3 H^2_0/2c^2=8.57 \times 10^{-53} m^{-2}$. \footnote{The factor $c^2$ could already have appeared in the right hand side of (\ref{cec}) but now we have restored the SI units via the convertion $\Lambda_{\rm U} \rightarrow \Lambda_{\rm U} c^2$.} By fixing $H_0=70 km/s/Mpc$ and $q_0=-0.5$ one can estimate from (\ref{q0}) the value $\Lambda_{\rm U} \cong 1.29 \times 10^{-52} m^{-2}$ which is of the same order of magnitude as the value obtained for the ``traditional'' cosmological constant $\Lambda$ in the concordance $\Lambda$CDM model.

The expansion rate written in terms of the scale factor reads
\begin{equation}
    H(a)=H_0 \left(\Omega_{\rm U}+\frac{1-\Omega_{\rm U}}{a^4}\right)^{1/2},
\label{HU}
\end{equation}
where we have defined the parameter
\begin{equation}
    \Omega_{\rm U}= \frac{\Lambda_{\rm U}}{3 H^2_0}.
    \label{OmegaU}
\end{equation}

Whereas the expansion rate presented in (\ref{HU}) resembles the GR case sourced by radiation and cosmological constant, the physical interpretation is different. The parameter $\Omega_{\rm U}$ is uniquely related to the constant $\Lambda_{\rm U}$ via (\ref{OmegaU}). All relativistic and non-relativistic species should sum up to the quantity $1-\Omega_{\rm U}$. Indeed, even non-relativistic matter will scale as $\sim a^{-4}$ in the nonconservative unimodular gravity (NUG).

Expansion rate (\ref{HU}) is indeed different from $\Lambda$CDM since it does not admit a matter dominated phase. Of course, a quantitative statistical analysis using current available data would disfavor expansion (\ref{HU}) in comparison to the standard $\Lambda$CDM model. However, let us investigate whether or not expansion (\ref{HU}) can be considered viable.

We start calculating the age of the universe $t_{univ}$ as a function of the parameters $H_0$ and $\Omega_{\rm U}$ via
\begin{equation}
    t_{univ} = \int^{1}_{0}\frac{da^{\prime}}{a^{\prime} H(a^{\prime})}.
\end{equation}

\begin{figure}[!t]
\includegraphics[width=\linewidth]{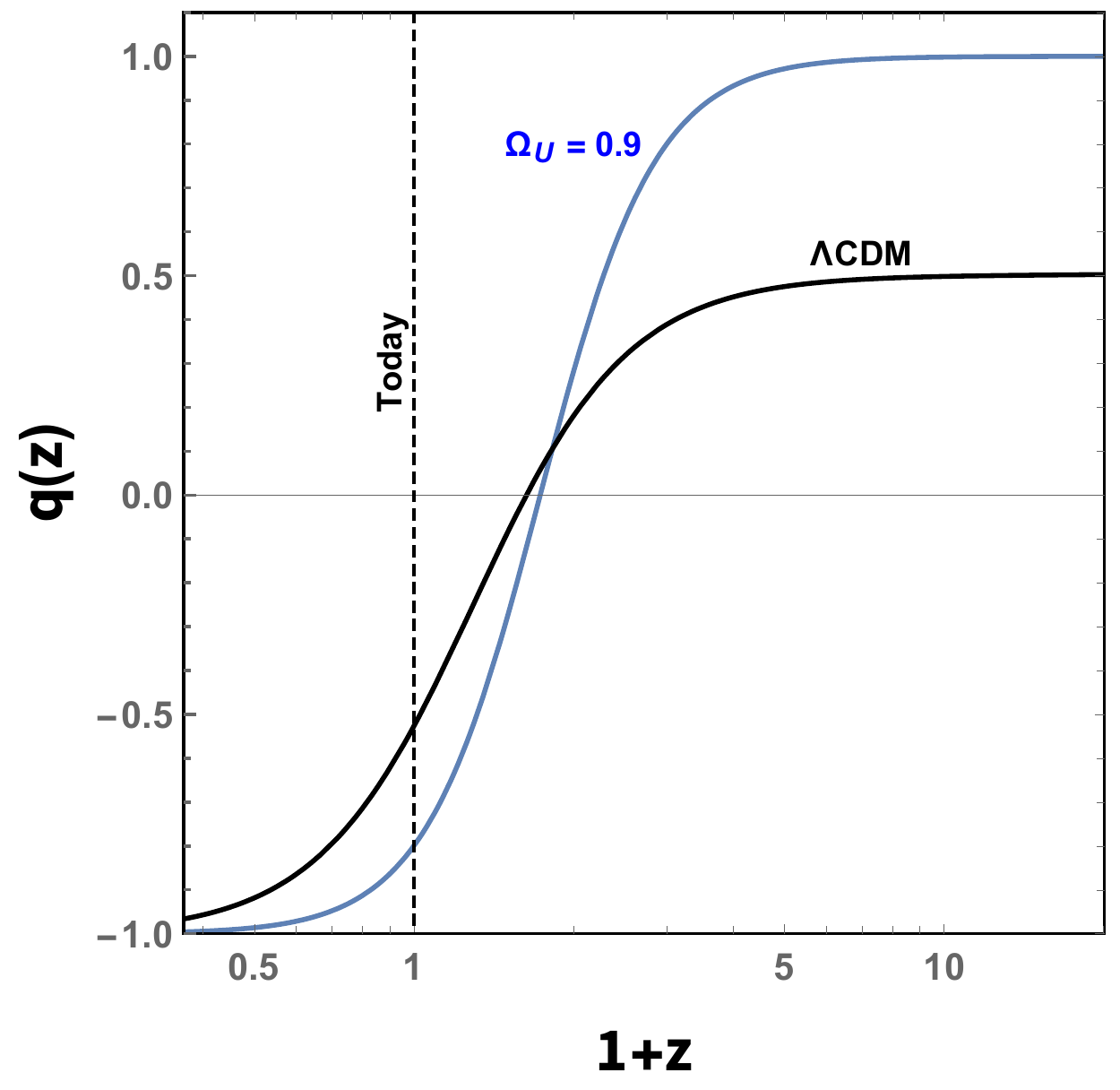}
\caption{Evolution of the deceleration parameter as a function of the redshift $z$. The vertical dashed line corresponds to the redshift $z=0$. }
	\label{figq}
\end{figure}

By fixing $H_0=67.3 km/s/Mpc$ and $\Omega_{\rm U}=0.9$ one finds for the age of the universe $13.9 \,Gyrs$ in agreement with standard cosmology estimations. The larger $\Omega_{\rm U}$ the older is the universe. This also means that the universe is older than the estimated ages of globular cluster which are considered the oldest known objects. Let us then now adopt such parameter values and plot the deceleration parameter as a function of the redshift $q(z)$. In Fig. \ref{figq} the evolution of the deceleration parameter is shown for the concordance $\Lambda$CDM model in the black line and for the NUG cosmology with $\Omega_{\rm U}=0.9$ in the blue line.  The latter transits from the radiative decelerated phase with $q=1$ to the accelerated one earlier than the $\Lambda$CDM model at the redshit $z_{tr}\sim 0.73$ reaching the today's deceleration parameter $q_0=-0.8$.

Concerning structure formation the absence of a matter dominated epoch is not accepted in GR but as we shall show in the next section scalar perturbations will behave differently and can potentially yield a viable scenario.

\section{Scalar Cosmological Perturbations in unimodular gravity}

The analysis of cosmological perturbations in the unimodular theory allows us to address to issue of fixing the coordinate condition.
In fact, by considering the metric
\begin{eqnarray}\label{pertmetric}
ds^2 = a^2\biggr\{(1 + 2\phi)d\eta^2 - 2 B_{,i}dx^id\eta \nonumber \\
- [(1 - 2\psi)\delta_{ij} + 2E_{,i,j}]dx^i dx^j\biggl\},
\end{eqnarray}
one can discuss the main options for the perturbative analysis: the synchronous coordinate condition ($\phi = B = 0$), the newtonian coordinate condition ($B = E = 0$) and the gauge invariant formalism. As discussed below, the gauge issue is a subtle aspect of unimodular gravity.
 
 By perturbing the unimodular condition with (\ref{pertmetric}) one finds,
 \begin{eqnarray}\
 \label{puc}
 \nabla^2E + \phi - 3\psi = 0.
 \end{eqnarray}
As already pointed out in \cite{Gao:2014nia}, the gauge freedom imposed by (\ref{puc}) is different from the GR case and both the newtonian gauge and the synchronous gauge do not apply in the unimodular theory. Let us review this.  It is clear from (\ref{puc}) that there is a problem with the newtonian coordinate condition since it implies,
\begin{eqnarray}
\phi - 3\psi = 0.
\end{eqnarray}
However, in the absence of anisotropic stress, the newtonian coordinate condition implies $\phi = \psi$. Hence, the only accepted solution is the trivial one i.e., $\phi=\psi=0$. Of course, in the presence of the anisotropic stress the situation is more involved and requires a detailed analysis. In what follows we will not consider anisotropic stress.

On the other hand, for the gauge invariant formalism the restriction (\ref{puc}) means that the unimodular condition implies in a restricted class of coordinate transformation.

Finally, using the synchronous coordinate condition, we can define the quantity,
\begin{eqnarray}
h = \frac{h_{kk}}{a^2} = 2(3\psi - \nabla^2E).
\end{eqnarray}
From (\ref{puc}), with $\phi = 0$, we obtain $h = 0$. If we impose the conservation of the energy-momentum tensor, the quantity $h$ is directly connected with matter perturbations. In fact, in the usual FLRW universe, the perturbed equation connecting
the function h and the matter perturbation can be written as,
\begin{equation}
    \ddot{h}+2H\dot{h}=4 \pi G (1+ 3 v^2_s)\delta \rho.
\end{equation}
with $v^2_s$ indicating the sound velocity. This equation is valid also in unimodular gravity if the conservation of the energy-momentum tensor is
preserved. But, if $h = 0$, due to the unimodular condition, than $\delta \rho =0$,
and no matter perturbation is present when the synchronous coordinate
condition is used.

Hence, there are no perturbations at all if this condition is chosen. 

One of the results presented in \cite{Gao:2014nia} determines that one can not realise scalar perturbations in the synchronous and longitudinal gauges in unimodular gravity. 

\section{Cosmological perturbations when $T^{\mu\nu}_{\quad ;\mu}\neq0$.}

We show now that if the usual energy-momentum tensor is not separately conserved the restriction to the longitudinal gauge persists whereas it is possible to use the synchronous coordinate condition.

We start showing that the longitudinal gauge continues to show a pathological behavior in the nonconservative unimodular gravity. The line element in this gauge has the form,
\begin{eqnarray}
ds^2 = (1 + 2\phi)dt^2 - a^2(1 + 2\psi)\delta_{ij}dx^i dx^j.
\end{eqnarray}

The perturbed components of the Ricci tensor and the perturbed Ricci scalar are,
\begin{equation}
\delta R_{00} = - 3\ddot\psi + 3H(\dot\phi - 2\dot\psi) + \frac{\nabla^2\phi}{a^2},\\
\end{equation}
\begin{equation}
\delta R_{0i} = - 2\partial_i(\dot\psi - H\phi),
\end{equation}
\begin{eqnarray}
\delta R_{ij} = a^2\delta_{ij}\biggr\{\ddot\psi + H(6\dot\psi - \dot\phi) + \nonumber \\ 2(\dot H + 3H^2)(\psi - \phi) - \frac{\nabla^2\psi}{a^2}\biggl\}
- \partial_i\partial_j(\psi + \phi),
\end{eqnarray}
\begin{equation}
    \delta R = - 6\ddot\psi - 6H (4\dot\psi - \dot\phi) + 12(\dot H + 2H^2)\phi + 2\frac{\nabla^2}{a^2}(2\psi + \phi).
\end{equation}
The perturbed components of the energy-momentum tensor and its trace are,
\begin{eqnarray}
\delta T_{00} &=& \delta\rho  + 2\phi\rho,\\
\delta T_{0i} &=& - a^2(\rho + p)\delta u^i,\\
\delta T_{ij} &=& a^2\delta_{ij}(\delta p + 2p\psi),\\
\delta T &=& \delta\rho - 3\delta p.
\end{eqnarray}

Perturbing the unimodular equations we have,
\begin{eqnarray}
\delta R_{\mu\nu} - \frac{1}{4}(g_{\mu\nu}\delta R + h_{\mu\nu}R) = \nonumber \\ 8\pi G\biggr\{\delta T_{\mu\nu} - \frac{1}{4}(g_{\mu\nu}\delta T + h_{\mu\nu}T)\biggl\}.
\end{eqnarray}
Considering all the above definitions and fixing $\mu = i, \nu = j$, we obtain the equation for the gravitational potentials $\phi$ and $\phi$,
\begin{eqnarray}
\delta_{ij}&\biggr\{&\ddot\psi - H\dot\phi + 2H(\psi - \phi) - \frac{\nabla^2\phi}{a^2}\biggl\} + \frac{2}{a^2}\partial_i\partial_j(\psi + \phi) \nonumber\\= &-& 4\pi G\delta_{ij}\biggr\{(\delta\rho + \delta p) + 2\psi(\rho + p)\biggl\}.
\end{eqnarray}
Considering $i \neq j$, we obtain,
\begin{eqnarray}
\partial_i\partial_j(\psi + \phi) = 0,
\end{eqnarray}
implying $\psi = - \phi$. This would not be the case if one considers, e.g., anisotropic stresses. Combined with the perturbed unimodular condition in the same gauge, it comes out that $\psi = \phi = 0$. The newtonian gauge can not be used in the unimodular context unless any anisotropic contribution to the stress-tensor are considered.

Let us consider now the synchronous gauge including small quantities around the background ones such that,
\begin{eqnarray}
\tilde g_{\mu\nu} &=& g_{\mu\nu} + h_{\mu\nu},\\
\tilde\rho &=& \rho + \delta\rho,\\
\tilde u^i &=& u^i + \delta u^i.
\end{eqnarray}
In these expressions the quantities with tildes are the full ones, without tildes the background ones, and those preceded of $\delta$ as well as $h_{\mu\nu}$ are the perturbed quantities.

We use the synchronous coordinate condition,
\begin{eqnarray}
h_{\mu0} = 0.
\end{eqnarray}
Now, we deviate from the approach followed in Ref. \cite{Daouda:2018kuo}, in the sense that we take into account the unimodular constraint.
The condition $g = $ constant implies, at perturbative level,
\begin{eqnarray}
h_{kk} = 0.
\end{eqnarray}
The components of the perturbed Ricci tensor and the Ricci scalar are given by:
\begin{eqnarray}
\delta R_{00} &=& 0,\\
\delta R_{0i} &=& - \frac{1}{2}\biggr(\frac{h_{ki,k}}{a^2}\biggl)^{.},\\
\delta R_{ij} &=& \frac{1}{2a^2}\biggr(\nabla^2 h_{ij} - h_{ki,j,k} - h_{kj,i,k}\biggl) - \frac{\ddot h_{ij}}{2} + \frac{H}{2}\dot h_{ij} \nonumber \\- 2 H^2h_{ij}, \label{deltaR3}\\
\delta R &=& \frac{h_{jk,j,k}}{a^4}.
\end{eqnarray}

The perturbations of the components of the energy-momentum tensor are,
\begin{eqnarray}
\delta T^{00} &=& \delta\rho,\\
\delta T^{0i} &=& (\rho + p)\delta u^i,\\
\delta T^{ij} &=& \frac{\delta p}{a^2}\delta_{ij} + h_{ij}\frac{p}{a^4},\\
\delta T &=& \delta\rho - 3\delta p.
\end{eqnarray}

We define the scalar metric perturbation $f$ and the velocity potential perturbation $\theta$  as,
\begin{equation}
f = \frac{h_{kj,k,j}}{a^2}, \quad
\theta = \delta u^i_{,i}.
\end{equation}

With the above definitions the perturbed field equations become 
\begin{eqnarray}
\label{pe1}
f &=& - 24\pi Ga^2(\delta\rho + \delta p),\\
\label{pe2}
\dot f &=& 16\pi G a^2(\rho + p)\theta.
\end{eqnarray}
From the conservation equations we obtain the following perturbed equations: \footnote{ This result is also obtained with the double divergence of (\ref{deltaR3}).}
\begin{eqnarray}
\label{pe3}
\dot f - 2Hf = 24\pi Ga^2\biggr\{\delta\dot\rho + \delta\dot p + 4H(\delta\rho + \delta p) + \nonumber \\ \frac{4}{3}(\rho + p)\theta\biggl\},\\
\label{pe4}
\frac{\nabla^2 f}{a^4} = - 32\pi G\biggr\{[(\rho + p)\theta]^. + 5H(\rho + p)\theta  \nonumber \\+ \frac{\nabla^2(\delta\rho + \delta p)}{4a^2}\biggl\}.\label{pe4}
\end{eqnarray}

Inserting (\ref{pe1}), (\ref{pe2}) into (\ref{pe4}) we obtain an identity i.e.,  equation (\ref{pe3}) contains no new information. However, the combination of (\ref{pe1}), (\ref{pe2}) and (\ref{pe3}) leads to,
\begin{eqnarray}
\label{pe}
\ddot f + 3H\dot f - \frac{k^2}{3a^2}f = 0.
\end{eqnarray}
In the above equation we have performed the Fourier expansion via the replacement $\nabla^2 \rightarrow - k^2$.

 At this point, we must fix the behavior of the scale factor in order to find an explicit solution. We will fix the case $\Lambda_{\rm U} = 0$ since we are interested in the evolution of the scalar perturbations starting at high redshifts until they reach the nonlinear stage.
The final solution in terms of the conformal time ($\eta \propto t^{1/2}$) reads
\begin{eqnarray}
f = A\frac{\sinh \frac{k}{\sqrt{3}}\eta}{k\eta} + B\frac{\cosh \frac{k}{\sqrt{3}}\eta} {k \eta}.
\label{fsolution}
\end{eqnarray}

Notice that there is asymptotically an exponential growth of the perturbations. Another important point is that the growing mode remains constant at large scales ($k \rightarrow 0$) and grows exponentially at very small scales ($k \rightarrow \infty$).

A finite solution at $\eta = 0 $ is obtained from (\ref{fsolution}) by restricting it to the $\sinh$ mode i.e., we set $B=0$.
Using the background solution for the $\Lambda=0$ case i.e., $a\sim \eta$ and the perturbative relations obtained above we can write the expression for the density contrast  
\begin{eqnarray}\label{deltabar}
\bar\delta = \frac{\delta \rho+\delta p}{\rho+p}= \bar A a^2\frac{\sinh q}{ q},
\end{eqnarray}
with $q = \frac{k a}{\sqrt{3}a_0}$ and $\bar{A}$ is a new constant redefined from $A$.
For large wavelengths, $k \rightarrow 0$, the density contrast behaves as,
\begin{eqnarray}
\bar\delta \sim a^2. 
\label{deltaa2}
\end{eqnarray}
This solution is valid even for the case of a pressureless fluid. Then, matter perturbations can grow even in a radiation-like expanding background. 

It is worth noting that in the GR case the pressureless matter perturbations grow as $\delta_{GR}\sim a$. Thus, the nonconservative unimodular case yields to a much faster growth than the corresponding relation during the matter dominated era in the Standard Cosmological Model.
This enhancement in the evolution of the matter perturbations is preserved for any wavelengths becoming even larger for small scales. This means that, contrarily to standard cosmological picture, matter perturbations can substantially grow during the radiation dominated phase which, in this model, extends up to the transition to the accelerated expansion.

In the late time regime, dominated by the cosmological constant, $H$ becomes constant and the perturbations stabilize. This can be directly inferred from Eq. (\ref{pe}).

Though (\ref{deltaa2}) is in clear contrast to standard matter perturbations growth it can indeed yield to the same amplitudes for today's matter density field if the initial amplitudes are different. For the $\Lambda$CDM model let us set the amplitude of typical galaxy cluster scales at the equality time $z_{eq}$ \footnote{The moment at which energy densities in matter and radiation are the same which in the $\Lambda$CDM model corresponds to a redshift $z_{eq}\sim 3400$.} as $\delta(z_{eq}) \sim 5 \times 10^{-4}$. If this perturbation evolves in time according to $\delta_{GR} \sim a$ they reach the nonlinear regime (i.e., $\delta_{GR}\sim 1$) recently in agreement with current large scale structure observations. On the other hand, in the nonconservative unimodular gravity with matter growth evolution given by (\ref{deltaa2}) it is also possible to reach the nonlinear regime if initial matter density contrast amplitudes are of order $\sim 10^{-7}$. Therefore, according to this estimation,  since the perturbations permitted in the nonconservative unimodular gravity are smaller than the ones we obtain in standard cosmology, a new mechanism to generate them should be investigated and compared to competitive inflationary models.

Finally, we have also verified that scalar perturbations in a gauge invariant formalism and found that the system is underdetermined. For the background dynamics our solutions are the same as in the case $R = cte$ as done in Ref. \cite{Daouda:2018kuo}. But once again we have not used the latter condition. Concerning the perturbations, if we had imposed $R = cte$ this would lead to a constrained theory with different perturbative results.

\section{Conclusions}

The issue of energy-momentum conservation in unimodular gravity is constantly discussed in the literature. In this work we have explored the consequences of evading the imposition of $T^{\mu}_{\nu;\mu}=0$ both at the level of an expanding cosmological background and its scalar perturbations. 

Rather than imposing the condition $R= cte $ as in Ref. \cite{Daouda:2018kuo} and even without any other constraint we have verified the existence of viable solutions.

Contrarily to the conservative case, in which the synchronous and longitudinal (Newtonian) gauges are not available, scalar perturbations in the nonconservative unimodular gravity are permitted in the synchronous gauge only and have a growing mode.

For any perfect fluid the background evolution behaves as a pure radiation dominated phase transiting to a future de Sitter epoch. Even in such non-standard scenario, the growth of scalar perturbations found here is potentially able to conduct primordial matter fluctuations to the non-linear regime of structure formation.

In order to make the scenario analyzed here unique, in view of the underdetermined aspect of the system of equations, it is necessary to impose further constraints on the
theory. One possibility is to implement it via an extra ingredient like the holographic principle.

Of course, one still have to verify whether such scenario remains viable after confrontation with observational data. We do not expect that the statistical analysis will provide very competitive results in comparison with the $\Lambda$CDM model. But interestingly, this result allows on to consider a non-standard cosmological model in which the matter dominated expansion epoch is absent.

\begin{acknowledgments}

The authors thank FAPES/CNPq/CAPES and Proppi/UFOP for financial support. We also thank Nelson Pinto-Neto for clarifying discussions.

\end{acknowledgments}

%

\end{document}